\input harvmac
\let\includefigures=\iftrue
\let\useblackboard=\iftrue
\newfam\black

\includefigures
\message{If you do not have epsf.tex (to include figures),}
\message{change the option at the top of the tex file.}
\input epsf
\def\figin{\epsfcheck\figin}\def\figins{\epsfcheck\figins}
\def\epsfcheck{\ifx\epsfbox\UnDeFiNeD
\message{(NO epsf.tex, FIGURES WILL BE IGNORED)}
\gdef\figin##1{\vskip2in}\gdef\figins##1{\hskip.5in}
\else\message{(FIGURES WILL BE INCLUDED)}%
\gdef\figin##1{##1}\gdef\figins##1{##1}\fi}
\def\DefWarn#1{}
\def\figinsert{\goodbreak\midinsert}
\def\ifig#1#2#3{\DefWarn#1\xdef#1{fig.~\the\figno}
\writedef{#1\leftbracket fig.\noexpand~\the\figno}%
\figinsert\figin{\centerline{#3}}\medskip\centerline{\vbox{
\baselineskip12pt\advance\hsize by -1truein
\noindent\footnotefont{\bf Fig.~\the\figno:} #2}}
\endinsert\global\advance\figno by1}
\else
\def\ifig#1#2#3{\xdef#1{fig.~\the\figno}
\writedef{#1\leftbracket fig.\noexpand~\the\figno}%
\global\advance\figno by1} \fi

\def\id{{1 \kern-.28em {\rm l}}}

\def\K3{{\bf K3}}
\def\journal#1&#2(#3){\unskip, \sl #1\ \bf #2 \rm(19#3) }
\def\andjournal#1&#2(#3){\sl #1~\bf #2 \rm (19#3) }

\def\hat{\widehat}
\def\ie{{\it i.e.}}
\def\eg{{\it e.g.}}

\def\tilde{\widetilde}

\def\frac#1#2{{#1\over#2}}

\def\inbar{\,\vrule height1.5ex width.4pt depth0pt}
\def\IC{\relax\hbox{$\inbar\kern-.3em{\rm C}$}}
\def\IR{\relax{\rm I\kern-.18em R}}
\def\IP{\relax{\rm I\kern-.18em P}}

%
%

%
\catcode`\@=11
\def\slash#1{\mathord{\mathpalette\c@ncel{#1}}}
\overfullrule=0pt

\def\eps{\epsilon}

\def\underrel#1\over#2{\mathrel{\mathop{\kern\z@#1}\limits_{#2}}}

\catcode`\@=12


%


\lref\GiveonZN{
  A.~Giveon and D.~Kutasov,
  ``Seiberg Duality in Chern-Simons Theory,''
  Nucl.\ Phys.\  B {\bf 812}, 1 (2009)
  [arXiv:0808.0360 [hep-th]].
}

\lref\IntriligatorDD{
  K.~A.~Intriligator, N.~Seiberg and D.~Shih,
  ``Dynamical SUSY breaking in meta-stable vacua,''
  JHEP {\bf 0604}, 021 (2006)
  [arXiv:hep-th/0602239].
}

\lref\GiveonEW{
  A.~Giveon and D.~Kutasov,
  ``Stable and Metastable Vacua in Brane Constructions of SQCD,''
  JHEP {\bf 0802}, 038 (2008)
  [arXiv:0710.1833 [hep-th]].
}
\lref\GiveonEF{
  A.~Giveon and D.~Kutasov,
  ``Stable and Metastable Vacua in SQCD,''
  Nucl.\ Phys.\  B {\bf 796}, 25 (2008)
  [arXiv:0710.0894 [hep-th]].
}
\lref\GiveonFK{
  A.~Giveon and D.~Kutasov,
  ``Gauge symmetry and supersymmetry breaking from intersecting branes,''
  Nucl.\ Phys.\  B {\bf 778}, 129 (2007)
  [arXiv:hep-th/0703135].
}
\lref\EssigKZ{
  R.~Essig, J.~F.~Fortin, K.~Sinha, G.~Torroba and M.~J.~Strassler,
  ``Metastable supersymmetry breaking and multitrace deformations of SQCD,''
  arXiv:0812.3213 [hep-th].
}

\lref\GiveonSR{
  A.~Giveon and D.~Kutasov,
  ``Brane dynamics and gauge theory,''
  Rev.\ Mod.\ Phys.\  {\bf 71}, 983 (1999)
  [arXiv:hep-th/9802067].
}

\lref\GiveonUR{
  A.~Giveon, D.~Kutasov, J.~McOrist and A.~B.~Royston,
  ``D-Term Supersymmetry Breaking from Branes,''
  arXiv:0904.0459 [hep-th].
}

\lref\WittenDS{
  E.~Witten,
  ``Supersymmetric index of three-dimensional gauge theory,''
  arXiv:hep-th/9903005.
}

\lref\NiarchosJB{
  V.~Niarchos,
  ``Seiberg Duality in Chern-Simons Theories with Fundamental and Adjoint
  Matter,''
  JHEP {\bf 0811}, 001 (2008)
  [arXiv:0808.2771 [hep-th]].
}

\lref\NiarchosAA{
  V.~Niarchos,
  ``R-charges, Chiral Rings and RG Flows in Supersymmetric Chern-Simons-Matter
  Theories,''
  arXiv:0903.0435 [hep-th].
}

\lref\AmaritiRB{
  A.~Amariti, D.~Forcella, L.~Girardello and A.~Mariotti,
  ``3D Seiberg-like Dualities and M2 Branes,''
  arXiv:0903.3222 [hep-th].
}

\lref\KitaoMF{
  T.~Kitao, K.~Ohta and N.~Ohta,
  ``Three-dimensional gauge dynamics from brane configurations with
  (p,q)-fivebrane,''
  Nucl.\ Phys.\  B {\bf 539}, 79 (1999)
  [arXiv:hep-th/9808111].
}

\lref\duncan{
M.~J.~Duncan and L.~G.~Jensen,
  ``Exact tunneling solutions in scalar field theory,''
  Phys.\ Lett.\  B {\bf 291}, 109 (1992).
}

\lref\siani{
  A.~Amariti and M.~Siani,
  ``R-symmetry and supersymmetry breaking in 3D WZ models,''
  JHEP {\bf 0908}, 055 (2009)
  [arXiv:0905.4725 [hep-th]].
}

\Title{}
{\vbox{\centerline{Spontaneous SUSY Breaking in Various Dimensions}
\bigskip
\centerline{}
}}
\bigskip

\centerline{\it Amit Giveon\foot{Permanent address: Racah
Institute of Physics, The Hebrew University, Jerusalem 91904, Israel.},
David Kutasov and Oleg Lunin}
\bigskip
\smallskip
\centerline{EFI and Department of Physics, University of
Chicago} \centerline{5640 S. Ellis Av., Chicago, IL 60637, USA }

\smallskip

\vglue .3cm

\bigskip

\let\includefigures=\iftrue
\bigskip
\noindent
We generalize the ISS model of spontaneous supersymmetry breaking to lower dimensions.
We also comment on the dynamics of the corresponding brane systems in string theory,
and on possible applications to gauge/gravity duality.

\bigskip

\Date{}

\newsec{Introduction}

In this note we will discuss a class of gauge theories in $d<4$ spacetime
dimensions that exhibit spontaneous supersymmetry breaking, either in the
vacuum or in metastable states. Our main motivation for studying these
theories comes from gauge/gravity duality. One way to study supersymmetry
breaking in $(d+1)$-dimensional gravity is via a $d$-dimensional gauge theory
dual. If the gravitational theory describes a large smooth $(d+1)$-dimensional
spacetime, the dual gauge theory is typically strongly coupled, and thus difficult
to solve. We will restrict attention to weakly coupled gauge theories, which are
easier to analyze, but  our results may be useful for studying gauge theories
with semiclassical gravity duals.

The models we will study can be introduced by starting with the ISS model
of spontaneous supersymmetry breaking in four dimensions \IntriligatorDD,
and dimensionally reducing it to $d<4$. This gives a gauge theory with four
supercharges, and it is natural to ask whether it breaks supersymmetry, like
its four dimensional analog. As in four dimensions, one can choose the
parameters of the model such that the gauge interactions are weak and can be
neglected. In that regime, the low energy degrees of freedom are chiral superfields,
and their dynamics is  described by a Wess-Zumino (WZ) model.  In four
dimensions, this model was studied by ISS; our main purpose is to generalize
their results to $d<4$. We will also discuss the regime of validity of the  field
theory analysis, and its relation to the dynamics of the brane system in string
theory that realizes this WZ model at low energies.

The case $d=3$ is particularly interesting for gauge/gravity duality, since the
corresponding gravitational theory is $3+1$ dimensional. In that case one can
replace the standard Yang-Mills (YM) kinetic term by a Chern-Simons (CS) one.
Many CS matter theories are expected to have $AdS_4$ holographic duals, and thus
provide a natural arena for connecting our discussion to four dimensional gravity.

This note is organized as follows. In section 2 we study the WZ models obtained by
reducing  the (generalized) ISS theory to $d<4$ dimensions. We show that they have
metastable supersymmetry breaking vacua, which can be studied reliably at weak
coupling. In section 3 we discuss the gauging of the color group, with either a YM or
(for $d=3$) CS kinetic term for the gauge field. In section 4 we summarize and
comment on our results, and the corresponding brane systems in string theory.

\newsec{SUSY breaking in $d<4$ WZ models}

In this section we consider a WZ model in $d$ spacetime dimensions, with chiral
superfields  $\Phi_i^j$,  $q^i_a$, $\tilde q_i^a$, where $i,j=1,\dots, N_f$,
$a=1,\dots, N$, and
\eqn\nnf{N_f>N~.}
As we will see later, when the condition \nnf\ is not satisfied, the model does not
break supersymmetry, even in metastable states.

The K\"ahler potential for all the fields is taken to be canonical.
The superpotential is given by
\eqn\EqOne{
W=hq\Phi\tilde q+h\Tr\left(\frac{1}{2}\eps\mu\Phi^2-\mu^2 \Phi\right)~.}
The couplings $h$, $\epsilon$, $\mu$ are in general complex. Two of them can
be taken to be real and positive by redefining $\Phi$, $\tilde q$, $q$. We will
sometimes take all three couplings to have this property, for simplicity. In
that case, the interesting dynamics occurs at real $\Phi$. It is easy to repeat
the discussion for the general case.

For $\epsilon=0$, the model \EqOne\ is obtained by a dimensional reduction of that
discussed in \IntriligatorDD. For $\epsilon\not=0$, it is similarly related to the model
studied in \refs{\GiveonEF\GiveonEW\EssigKZ-\GiveonUR}. As we will see  later, a
non-zero $\epsilon$ is needed for $d<4$ to control the dynamics. The model \EqOne\
has a global symmetry $U(N)\times U(N_f)$. We will eventually be interested in gauging
$U(N)$,  but for now we treat it as a global symmetry.

The couplings that appear in the superpotential \EqOne\ have the following
mass dimensions:
\eqn\massdim{[h]=2-{d\over2}~,\qquad [\mu]={d\over2}-1~,\qquad [\epsilon]=0~.}
For $d<4$, all terms in the superpotential are relevant. Thus, in general
the WZ model \EqOne\ is expected to be strongly coupled in the infrared.
We will see later that this can be avoided when the vacua of interest are
located sufficiently far from the origin of field space.

The tree level bosonic potential corresponding to \EqOne\ is
\eqn\EqTri{V_0=|h|^2\left(|q^i{\tilde q}_j-\mu^2\delta^i_j+\eps\mu\Phi^i_j|^2+
|q\Phi|^2+|\Phi{\tilde q}|^2\right)~.}
Supersymmetric vacua are labeled by an integer $k=0,\cdots, N$. For given $k$,
the expectation values of the chiral superfields take the form (up to global symmetries)
\eqn\susyvac{
 \Phi=\left(\matrix{0&0\cr 0&\frac{\mu}{\epsilon}I_{N_f-k}}\right)~,\qquad
 q\tilde q =\left(\matrix{\mu^2I_k&0\cr 0&0}\right)~,}
where $I_k$ is a $k\times k$ identity matrix. In a vacuum with given $k$, the
$U(N)\times U(N_f)$ symmetry is broken to $U(N-k)\times U(N_f-k)\times U(k)$ by
the expectation values of $\Phi$, $q$, $\tilde q$.

In four dimensions, in addition to the supersymmetric vacua \susyvac, the WZ
model has metastable vacua in which the classical potential \EqTri\ is balanced
by a one loop, Coleman-Weinberg (CW) contribution \refs{\GiveonEF,\EssigKZ}.
It is natural to ask whether such vacua occur in lower dimensions as well.

Following  \refs{\GiveonEF\GiveonEW\EssigKZ-\GiveonUR}, we parameterize the fields
as follows:
\eqn\susybreakvac{
\Phi=\left(\matrix{0&0&0\cr 0&XI_n&0\cr 0&0&\frac{\mu}{\epsilon}I_{N_f-k-n}}\right)~,\qquad
q\tilde q =\left(\matrix{\mu^2I_k&0&0\cr 0&0&0\cr 0&0&0}\right)~,
 }
and examine the effective potential for X. The tree level potential \EqTri\  is
\eqn\EqFiv{V_0(X)=n|h\mu|^2|\mu - \eps X|^2.}
The Coleman-Weinberg potential is obtained by integrating out the components of
$\Phi$, $q$, $\tilde q$, whose mass depends on $X$. It is given by
\eqn\EqSix{V_1(X)=2nk\sum_{\eta,\sigma=\pm1}\left[ F_d(m_b)-F_d(m_f)\right]
+2n(N-k)\sum_{\eta=\pm1}\left[F_d(\hat m_b)-F_d(\hat m_f)\right]~,}
where
\eqn\vmm{F_d(m)=\frac12\int{d^dp\over (2\pi)^d}\ln(p^2+m^2)=
{|m|\over 4}~,\,-{m^2\over 8\pi}\ln m^2~,\,-{|m|^3\over 12\pi}~,\, {m^4\over 64\pi^2}\ln m^2~,}
in $d=1,2,3,4$, respectively.

The first term in \EqSix\ (proportional to $kn$) is the contribution of the $k\times n$ 
off-diagonal components of $q$, $\tilde q$ and $\Phi$ that transform in the fundamental 
of the $U(k)$ subgroup of $U(N)\times U(N_f)$ unbroken by the configuration  \susybreakvac. 
The second is due  to the components of $q$, $\tilde q$ that transform in the fundamental 
of the unbroken $U(N-k)$ subgroup. The masses in \EqSix\ depend on $X$ as follows \EssigKZ:
\eqn\eeqs{\eqalign{m_b^2=&|h^2|\left(|\mu|^2+\frac{1}{2}|X|^2+\frac{|\eps\mu|^2}{2}+\frac{\eta G}{2}+
\frac{\sigma}{2}\sqrt{(|X|^2-|\eps\mu|^2+\eta G)^2+4|\mu X^*+\eps|\mu|^2|^2}\right)~,
\cr
m_f^2=&|h^2|\left(|\mu|^2+\frac{1}{2}|X|^2+\frac{|\eps\mu|^2}{2}+
\frac{\sigma}{2}\sqrt{(|X|^2-|\eps\mu|^2)^2+4|\mu X^*+\eps|\mu|^2|^2}\right)~,\cr
\hat m_b^2=&|h|^2(|X|^2+\eta G)~,\quad G\equiv |\mu^2-\eps\mu X|~,\cr
\hat m_f^2=&|hX|^2~.\cr}}
The bosonic masses $m_b$ on the first line of \eeqs\ in the sector with $\eta=\sigma=-1$
are tachyonic near the origin,\foot{The masses $\hat m_b$ on the third line are tachyonic as
well for $\eta=-1$ and small $X$; we will return to them later.} and need to be treated more
carefully. One can think of the field configuration \susybreakvac\ as a deformation of the
supersymmetric vacuum, \susyvac, with $k\to k+n$. If the effective potential  for $q$, $\tilde q$,
$X$ has a non-supersymmetric local minimum of the form \susybreakvac, the global symmetry 
of the relevant sector of the theory is broken:
$U(k+n)\to U(k)\times U(n)$. This leads to $2kn$ Nambu-Goldstone bosons parametrizing
the coset $U(k+n)/U(k)\times U(n)$. These are precisely the $2kn$ modes with
$\eta=\sigma=-1$ mentioned above. Their masses must receive contributions due to one
loop effects that make them zero at the local minimum.

To analyze the one loop effective potential \EqSix, consider first the case $k=N$, for which the
second term in \EqSix\ is absent (\ie\ the $U(N)$ symmetry is completely broken). Expanding the
first term in \EqSix\ to second order in $X/\mu$ one finds
\eqn\voneloop{\eqalign{
d=1: & \qquad V_1=\frac{knh}{8}\left[4\mu(\sqrt{2}-2)+(4+\sqrt{2})\eps X+
2(\sqrt{2}-1)\frac{X^2}{\mu}+\cdots\right]~,\cr
d=2: & \qquad V_1=-\frac{knh^2}{4\pi}\left[\eps \mu X\ln(h^2\mu^2)+
(2X^2+2\mu^2+\eps\mu X)\ln 2-3\eps\mu X-2X^2+\cdots\right]~,\cr
d=3: & \qquad V_1=\frac{knh^3}{4\pi}\left[-\frac{4}{3}(\sqrt{2}-1)\mu^3+
(2-\sqrt{2})\eps\mu^2 X+(3-2\sqrt{2})\mu X^2+\cdots\right]~,\cr
d=4: & \qquad V_1=\frac{knh^4}{16\pi^2}\left[
\mu^3(\mu-2\eps X)\ln (h^2\mu^2)+
(\mu^4+\eps\mu^3 X+2\mu^2 X^2)\ln 4\right.\cr
&\qquad\qquad\qquad\qquad\left.-5\eps\mu^3 X-
2\mu^2 X^2+\cdots\right]~.\cr}}
The contributions from higher orders in the loop expansion and from nonperturbative effects can be neglected when
the coupling $h$ at the scale of the masses of the fundamentals $q$, $\tilde q$ is
small:
\eqn\smallmass{kh^2\ll \left(hX\right)^{4-d}~.}
The potential $V_0+V_1$ has a local minimum at\foot{Here we assume that $\eps^2\ll bk$, 
$|a|k\ll 1$; we will justify these approximations shortly.}
\eqn\EqEght{X_0\simeq\frac{\eps\mu(1+ak)}{\eps^2+bk}\approx {\eps\mu\over bk}~,}
where $a$ and $b$ are dimension dependent constants:
\bigskip
\vbox{
$$\vbox{\offinterlineskip
\hrule height 1.1pt
\halign{&\vrule width 1.1pt#
&\strut\quad#\hfil\quad&
\vrule width 1.1pt#
&\strut\quad#\hfil\quad&
\vrule width 1.1pt#
&\strut\quad#\hfil\quad&
\vrule width 1.1pt#\cr
height3pt
&\omit&
&\omit&
&\omit&
\cr
&\hfil $d$&
&\hfil $a$&
&\hfil $b$&
\cr
height3pt
&\omit&
&\omit&
&\omit&
\cr
\noalign{\hrule height 1.1pt}
height3pt
&\omit&
&\omit&
&\omit&
\cr
&\hfil $1$&
&\hfil $-\frac{4+\sqrt{2}}{16h\mu^3}$&
&\hfil $\frac{\sqrt{2}-1}{4h\mu^3}$&
\cr
height3pt
&\omit&
&\omit&
&\omit&
\cr
\noalign{\hrule}
height3pt
&\omit&
&\omit&
&\omit&
\cr
&\hfil $2$&
&\hfil $\frac{\ln(2h^2\mu^2)-3}{8\pi\mu^2}$&
&\hfil $\frac{1-\ln 2}{2\pi\mu^2}$&
\cr
height3pt
&\omit&
&\omit&
&\omit&
\cr
\noalign{\hrule}
height3pt
&\omit&
&\omit&
&\omit&
\cr
&\hfil $3$&
&\hfil $-\frac{(2-\sqrt{2})h}{8\pi\mu}$&
&\hfil $\frac{(3-2\sqrt{2})h}{4\pi\mu}$&
\cr
height3pt
&\omit&
&\omit&
&\omit&
\cr
\noalign{\hrule}
height3pt
&\omit&
&\omit&
&\omit&
\cr
&\hfil $4$&
&\hfil $\frac{h^2}{32\pi^2}\left(5+2\ln\frac{h^2\mu^2}{2}\right)$&
&\hfil $\frac{\ln 4-1}{8\pi^2}h^2$&
\cr
height3pt
&\omit&
&\omit&
&\omit&
\cr
}\hrule height 1.1pt
}
$$
}
\noindent
The expression for the position of the local minimum, \EqEght, is valid when $\epsilon$ is in the
range\foot{Here and below we restrict to $d<4$.}
\eqn\condeps{bk\gg\epsilon\gg bk{k^{1\over 4-d}h^{d-2\over 4-d}\over\mu}~,}
where the lower bound comes from the condition for validity of the one loop approximation,
\smallmass, and the upper bound is  the condition $X\ll\mu$, which was used in approximating
the Coleman-Weinberg potential by \voneloop.

Note that the existence of the range \condeps\ implies a lower bound on $\mu$:
\eqn\lowmu{\mu^{4-d}\gg kh^{d-2}~.}
In this range,  $|a|k, bk,\eps \ll1$; this, together with \condeps, provides a justification for the 
approximations mentioned in footnote 3. To estimate the lifetime of the 
metastable vacuum, we need to compute the Euclidean action which gives the amplitude of tunneling to the true minimum. Approximating the barrier by a triangular potential and applying the arguments of 
\duncan\ 
to the $d$--dimensional field theory, we find\foot{The case of $d=3$ is discussed in more detail in \siani.}
\eqn\lifetime{
S\sim V^{1-d/2}_{\rm met}(X_{\rm susy})^d\sim (h\mu^2)^{2-d}\left(\frac{\mu}{\eps}\right)^d\gg \eps^{-d}~,
}
where in the last inequality we used \lowmu. 
For $\eps\ll1$, the non-supersymmetric state at 
\EqEght\ is parametrically long-lived, and is at a value of $X$ which is well separated from the 
supersymmetric vacuum, located at 
$X_{\rm susy}=\mu/\epsilon\gg\mu$.

We end this section with some comments:
\item{(a)} Validity of the loop expansion in the WZ model requires the one loop potential $V_1$
to be much smaller than the tree level contribution $V_0$. Equations \EqTri, \voneloop\ imply that
\eqn\vonesmall{{V _1\over V_0}\sim k{(h\mu)^d\over (h\mu^2)^2}=k{h^{d-2}\over \mu^{4-d}}\ll 1~.}
The last inequality in \vonesmall\ follows from \lowmu.
\item{(b)} At the local minimum \EqEght, the pseudo-moduli $X$ have a mass, which can be read
off from \voneloop,
\eqn\masspseudo{m_X^2=bk(h\mu)^2~.}
Since $bk$ is small in the regime \lowmu, the mass \masspseudo\ is well below the masses of the 
quarks \eeqs, which were integrated out in calculating the potentials \voneloop. Thus, the above analysis
is consistent.
\item{(c)} We restricted the discussion to the small field region, where the one-loop potential
can be approximated by \voneloop, and the effective coupling $h$ is small, \smallmass. It is
natural to expect that the results can be extended beyond this region. For example, as
$\epsilon\to0$, the supersymmetric vacuum goes to infinity, and the non-supersymmetric one,
\EqEght, approaches the origin.  The lower bound \condeps\ is violated, and the low energy
dynamics becomes strongly coupled. Although this limit is not under control, we expect the
theory with $\epsilon=0$ to have a stable, non-supersymmetric vacuum at the origin \GiveonZN.
\item{(d)} For $k<N$, some of the components of $q$, $\tilde q$ are tachyonic for small $X$
(see third line of \eeqs). To avoid these tachyons one must go to $X\sim\mu$, or larger.
Studying the one-loop corrected potential $V_0+V_1$ numerically, we find that, as in the four
dimensional analysis of \EssigKZ, there are no local minima of the full potential,
which do not decay by condensation of the above tachyonic modes.

\newsec{Gauging $U(N)$}

Unlike its four dimensional analog, in $d\le 3$ the WZ model  with superpotential \EqOne\
is UV complete. However, for applications to gauge/gravity duality it is interesting to consider
a generalization of the model in which part or all of the global symmetry group $U(N)\times
U(N_f)$ is gauged.

Consider, for example, the case where we gauge $U(N)$ (of course, requiring that the Lagrangian
still preserves SUSY).  This introduces into the problem a new parameter, the gauge coupling $g_{YM}$,
whose mass dimension is $[g_{YM}]=2-d/2$, the same as that of $h$, \massdim. The gauge interaction
is a relevant  perturbation of the WZ model, which grows in the infrared, just like $h$. The results
of the previous section are not modified by its presence when $g_{YM}\ll h$, so that the effect of
gauge theory loops is smaller than that of loops in the WZ model, taken into account in section 2.

The $U(N)$ gauge field belongs to a supermultiplet that also contains $4-d$ real scalar fields in
the adjoint representation. In analyzing the non-supersymmetric vacuum structure, one needs to 
determine the potential of these fields as well as of those considered in section 2. However, since 
in the WZ model SUSY breaking vacua occur only in  the sector with $k=N$, where the $U(N)$ 
symmetry is completely broken,  these scalars get a mass of order $g_{YM}\mu$, and do not 
destabilize the vacua found in section 2.

In three dimensions, one can also consider a theory in which the kinetic term of the $U(N)$ gauge
field takes the Chern-Simons form.  The Chern-Simons level $k_{cs}$ is quantized, but at large level
one can study the gauge dynamics perturbatively in the 't~Hooft coupling $N/k_{cs}$.  The analysis
of the previous section is valid to leading order in the 't~Hooft coupling.

In \GiveonZN\ it was argued that the above CS theory is related by a Seiberg-type duality\foot{For
generalizations of this duality, see \eg\ \refs{\NiarchosJB,\NiarchosAA,\AmaritiRB}.}
to $U(N_c)$ Chern-Simons theory at level $k_{cs}$, coupled to $N_f$ fundamentals $Q_i$, $\tilde Q^i$,
with superpotential
\eqn\electricw{W_{\rm el}=m\tilde Q Q+\lambda (\tilde Q Q)^2~.}
The parameters $\epsilon$ and $\mu$ in \EqOne\ are related to $m$ and $\lambda$,
while the field $\Phi$ is proportional to the electric meson field $\tilde Q Q$.
The rank of the magnetic gauge group $N$ is given by \GiveonZN:
\eqn\rankmm{N=N_f+k_{cs}-N_c~.}
The condition for supersymmetry breaking \nnf\ is  $k_{cs}<N_c$. It is believed
\WittenDS\ that for $\lambda=0$ the electric theory spontaneously breaks supersymmetry
for $k_{cs}$ in this range. This is consistent with the fact that the magnetic theory
is expected to break SUSY for $\epsilon=0$.

\newsec{Discussion}

The main conclusion of our analysis is that the appearance of spontaneous supersymmetry
breaking metastable vacua, found by ISS \IntriligatorDD\ in four dimensional gauge theory,
persists to lower dimensions as well. For generic values of the parameters, the $d<4$
dimensional gauge theory with superpotential \EqOne\ is strongly coupled in the
infrared, and the existence of such vacua is difficult to establish. However, there are
regions in parameter space where these vacua can be studied reliably at weak coupling.
The phase structure we found is similar to that found in four dimensions in
\refs{\IntriligatorDD,\GiveonEF,\EssigKZ}. For example, in terms of the decomposition
\susybreakvac, the branch with $k=N$ contains SUSY breaking vacua, while the
``tachyonic branches'' with $k<N$ do not.

As mentioned in the introduction, one of the primary motivations for this study is
gauge/gravity duality. A useful step in that direction is to realize the systems
we studied in string theory. This can be done by using the brane constructions
reviewed in \GiveonSR.

\ifig\loc{The brane configuration.}
{\epsfxsize5.0in\epsfbox{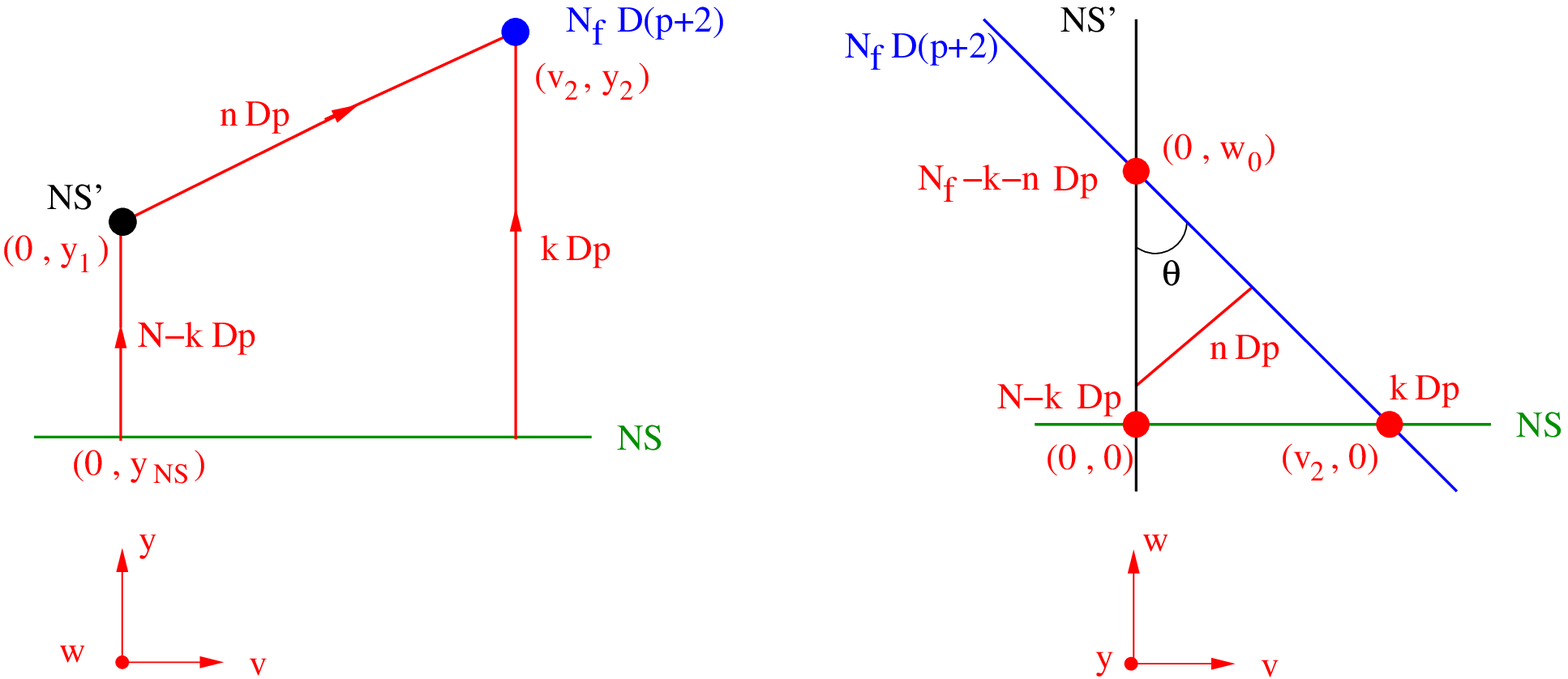}}

A brane system that realizes the gauge theory described in section 3 is exhibited in figure 1.
The $d$ dimensional gauge theory is realized on $Dp$-branes with $p=d$, stretched between
$NS5$-branes and other $D$-branes. The parameters of the WZ model \EqOne\ are related to
the brane ones as follows:
\eqn\parmatch{h^2={2g_s(2\pi)^{p-2}l_s^{p-3}\over y_2-y_1}~, \qquad
\mu^2={v_2\over2g_s(2\pi l_s)^{p-1}}~,\qquad \epsilon^2={2\pi l_s^2\tan^2\theta\over v_2(y_2-y_1)}~.}
Figure 1 gives rise to a gauge theory with standard kinetic term for the $U(N)$ gauge field.
As discussed in section 3, in three dimensions (\ie\ for $p=3$ in figure 1), one can also
consider a theory in which the kinetic term has a CS form. In the brane picture this corresponds
to replacing the $NS'$-brane by a $(1,k_{cs})$ brane, oriented such that supersymmetry is
unbroken \KitaoMF.

The relation between the low energy dynamics of the branes and the gauge theory in
four dimensions has been recently discussed in \GiveonUR. The picture in $d<4$ is
very similar. In the region of parameter space where the geometric description of figure
1 is reliable (when all the distances are much larger than $l_s$, and the string coupling
$g_s$ is small), the loop effects in the WZ model, \voneloop, and the corresponding
gauge theory effects, are small. The dominant corrections are due to the gravitational
attraction of the $Dp$-branes to the $NS$-brane in figure 1. The balance of this attraction
and the classical effects leads to the appearance of metastable vacua, which can be
studied as in \refs{\GiveonFK,\GiveonEW}. 

{}From the point of view of the low energy theory, these gravitational effects give rise to a 
non-trivial K\"ahler potential and higher D-terms for the pseudo-moduli, which together with 
the non-trivial superpotential stabilize them \GiveonUR. The brane picture leads to a phase structure 
which is essentially the same in all dimensions, since the D-terms that govern it are independent 
of $d$ when written in terms of geometric variables. Therefore, it is nice that the field theory 
analysis also gives a phase structure which is similar for different dimensions. 

Low energy theories on branes play an important role in gauge/gravity duality. It
would be interesting to apply our results in that context. A particularly promising
class of theories are three dimensional CS theories which often have supergravity
duals. We have seen that for large CS level, $k_{cs}\gg N$, the appearance of
supersymmetry breaking vacua is rather generic. It would be interesting to see whether
this is also the case in the gravity regime $N\gg k_{cs}\gg 1$.

\smallskip
\noindent{\bf Acknowledgements:} We thank G. Torroba for correspondence. This work is
supported in part by the BSF -- American-Israel Bi-National Science Foundation. AG is
supported in part by a center of excellence supported by the Israel Science Foundation
(grant number 1468/06), DIP grant H.52, and the Einstein Center at the Hebrew University.
DK and OL are supported in part by DOE grant DE-FG02-90ER40560 and the National Science
Foundation under Grant 0529954. AG thanks the EFI at the University of Chicago for
hospitality during the course of this work.

\listrefs
\end